\documentclass[preprint,12pt]{elsarticle}
\usepackage{amsmath,amssymb,amsfonts}
\usepackage{graphicx}
\usepackage{tabularx}
\usepackage{booktabs}
\usepackage{multirow}
\usepackage{float}
\usepackage{hyperref}
\usepackage{algorithm}
\usepackage{algorithmic}
\usepackage{xcolor}
\usepackage{setspace}
\usepackage{comment}
\usepackage{enumitem}
\usepackage{lineno}

\journal{Renewable and Sustainable Energy Reviews}

\usepackage{CJKutf8}

\begin{document}

\emergencystretch=1em

\begin{frontmatter}
%TC:ignore
% \title{Integrated survey of EV charging planning, scheduling, and behavior}
\title{Planning, Scheduling, and Behavior in EV Charging Systems: A Critical Survey and Trilemma Framework}

\author[wm]{Peiyan Xiao}
\ead{pxiao@wm.edu}

\author[wm]{Yuheng Li}
\ead{yli95@wm.edu}

\author[wm]{Ayan Mukhopadhyay}
\ead{amukhopadhyay@wm.edu}

\author[uga]{Sai Krishna Ghanta}
\ead{sai.krishna@uga.edu}

\author[louisville]{Sabur Baidya}
\ead{sabur.baidya@louisville.edu}

\author[wm]{Yanhai Xiong\corref{cor1}}
\ead{yanhaixiong7@gmail.com}

\cortext[cor1]{Corresponding author}

\address[wm]{The College of William \& Mary, Williamsburg, VA, USA}
\address[uga]{University of Georgia, Athens, GA, USA}
\address[louisville]{University of Louisville, Louisville, KY, USA}
%TC:endignore

\begin{abstract}
The rapid growth of electric vehicles is shifting the main constraint on transport electrification from vehicle adoption to the deployment and operation of charging infrastructure. Charging-network design requires decisions across three interdependent layers: Planning, which determines where and how much infrastructure to build; Scheduling, which governs charging dispatch, pricing, and grid interaction; and Behavior, which captures how users choose stations, charging times, and charging durations. Existing studies have advanced each layer substantially, but the literature remains fragmented, and cross-layer interactions are often treated through simplifying assumptions. This survey develops a three-layer Planning–Scheduling–Behavior (PSB) framework to organize EV charging research according to decision horizon, actor objective, and coupling structure. We further identify a fidelity–tractability tradeoff, termed the PSB trilemma: each layer is computationally difficult in isolation, and realistic integration across layers generally requires reducing the fidelity of at least one layer. Reviewing the three pairwise-coupling literatures—Planning–Scheduling, Scheduling–Behavior, and Planning–Behavior—we show that the omitted third layer is typically fixed exogenously or represented by a static aggregate surrogate. These simplifications enable tractability but impose distinct costs: they can obscure long-term investment feedback, temporal grid and emissions dynamics, or heterogeneous user response and equity outcomes. Building on this diagnosis, we identify open challenges in emerging charging technologies, behavioral incentives, equity metrics, and city-scale learning-based methods that balance fidelity, interpretability, and policy relevance.

\medskip
\noindent\textbf{Word count:} 9,881

\end{abstract}

%TC:ignore
%TC:endignore
\end{frontmatter}

\doublespacing

\section{Introduction}
% =====================================================
Achieving net-zero emissions requires deep decarbonization of transport, which produces about one-quarter of global energy-related CO$_2$ emissions; road vehicles account for nearly three-quarters of that share~\cite{Jaramillo2022Transport}. Because transport emissions are dispersed and shaped by long vehicle lifetimes and heterogeneous travel patterns, passenger and light-duty vehicle electrification is viewed as a key near-term mitigation pathway~\cite{Koroma2026Aligning}. With more than sixty countries adopting combustion-engine phase-out targets or EV mandates, global electric car stock reached nearly 58 million in 2024 and is projected to exceed 250 million by 2030~\cite{IEA2025GlobalEVOutlook}. As deployment accelerates, the main constraint is shifting from EV availability to charging infrastructure.

Charging-network deployment requires integrated decisions on long-term siting and capacity, short-term pricing and power dispatch, and user responses to price, wait time, detour, and availability. These decisions involve different actors and time scales but are tightly coupled: infrastructure constrains operations, operations shape charging choices, and aggregate demand determines utilization and investment viability. Yet the literature has largely developed each layer separately.

Planning studies have advanced from fixed-demand coverage models~\cite{WANG201376,KUBY2005125,church1974maximal} to queueing, grid-coupled sizing~\cite{motlagh2026_evcs_planning_review,mak2013infrastructure}, and multi-period expansion under uncertainty~\cite{zhang2023multiperiod,cui2025multistage}. Scheduling has moved from deterministic station dispatch~\cite{Sortomme2011OptimalCharging} to V2G, renewables, and multi-vehicle coordination~\cite{Motlagh2025EVCSOperationReview}, then to predictive and learning-based control under stochastic arrivals and prices~\cite{Hermans2024MPCVehicleCharging,Qiu2023RLEVReview}. Behavior models have progressed from exogenous demand~\cite{DCZHANG2015111,XI201360} to choice, equilibrium, and heterogeneous adaptive populations~\cite{HUANG2020102179,HE2014306,YI2023101949,yang2024latentclass}. Thus, each layer has matured in isolation, motivating cross-layer integration.

Studies that jointly address two or more stages have begun to appear. Planning models incorporate operational dispatch into the investment decision~\cite{prakash2023bilevel,tostadoveliz2024}; scheduling frameworks anticipate user response to pricing signals~\cite{zheng2025stackelberg,CHENG2025125145,BAE2024104540}; and siting models account for user choice behavior through equilibrium-based formulations~\cite{QIAO2023104678}. However, these cross-cutting studies remain comparatively limited, and most achieve tractability only by simplifying one stage to a fixed assumption. Existing surveys mirror this pattern: some integrate planning with power-distribution coupling but do not engage scheduling or user response~\cite{unterluggauer2022}; the most recent planning review does not address operational and behavioral feedback~\cite{motlagh2026_evcs_planning_review}; others focus on operations~\cite{Motlagh2025EVCSOperationReview,Sadeghian2022EVSmartChargingReview}, deterministic scheduling~\cite{DOLGUI2025221}, reinforcement-learning methods~\cite{Qiu2023RLEVReview}, or the service and policy landscape~\cite{lamonaca2022_state_of_play}.

This pattern is not accidental. Each of the three decision layers is individually a hard computational problem: facility location with integer variables is NP-hard, scheduling under stochastic arrivals and grid constraints leads to MDPs that suffer the curse of dimensionality, and computing user equilibria scales poorly with the number of user classes and alternatives. Coupling any two layers compounds the difficulty, producing problem classes (bilevel programs, stochastic games, mathematical programs with equilibrium constraints) for which no general-purpose efficient algorithms exist. The result is a tradeoff between modeling fidelity and computational tractability: practical joint formulations at a realistic scale generally simplify at least one layer. We call this the Planning-Scheduling-Behavior (PSB) trilemma.  

\begin{figure}[H]
    \centering
    \label{fig:intro-framework}
    \includegraphics[
        width=\linewidth,
        trim=200 100 280 100,
        clip
    ]{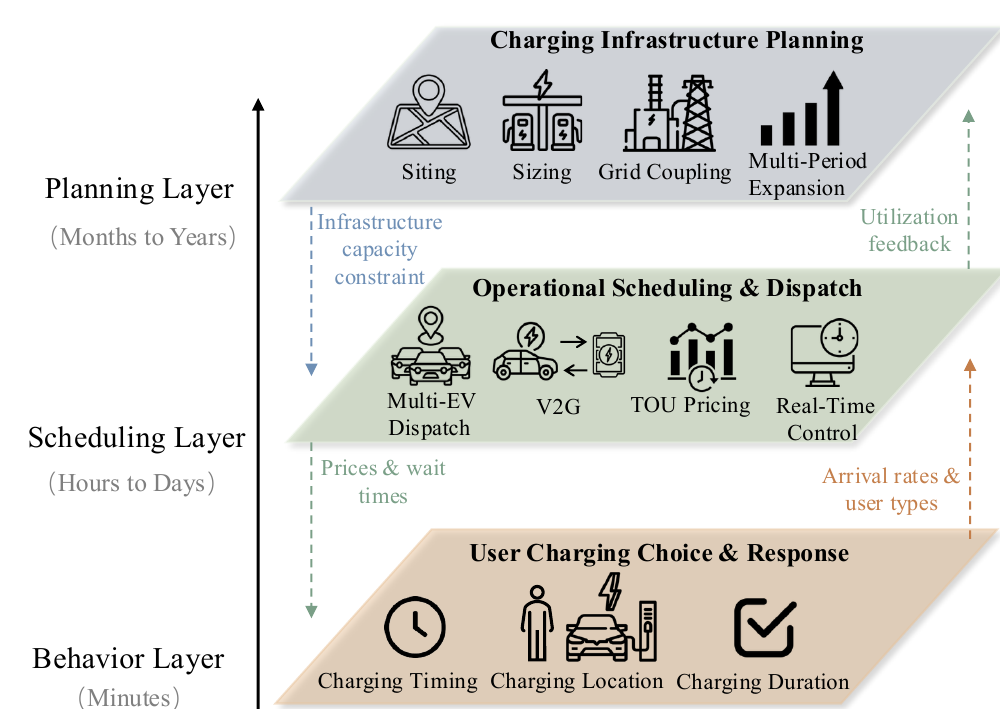}
    %TC:ignore
    \caption{Planning–Scheduling–Behavior framework for EV charging systems. The vertical axis indicates the decision horizon. Solid blocks denote the three decision layers; arrows denote coupling variables exchanged across layers. Planning constrains scheduling through infrastructure capacity; scheduling affects behavior through prices and waiting times; behavior feeds back through arrival rates, user types, utilization, and long-run demand.}
    %TC:endignore
\end{figure}

Existing reviews typically organize EV charging research by application or method; this survey instead uses cross-layer decision coupling. Its central claim is that fragmentation reflects not only a literature gap but also a fidelity--tractability tradeoff. By unifying Planning, Scheduling, and Behavior under the PSB framework, the survey makes each stream's assumptions, simplifications, and blind spots explicit, linking modeling choices to the practical needs of infrastructure deployment. Figure~\ref{fig:intro-framework} illustrates the three components and their information flows. We make three contributions:

\begin{enumerate}
    \item \textbf{A three-layer PSB framework with an actor taxonomy} (Section~\ref{sec:framework}). We define Planning, Scheduling, and Behavior by decision horizon, specify their coupling edges, and identify the actors and objectives associated with each layer. This allows studies to be compared by the layers they model, the couplings they retain, and the simplifications they impose.

    \item \textbf{The PSB trilemma} (Section~\ref{subsec:trilemma}). We identify a fidelity--tractability tradeoff caused by the computational difficulty of each layer and the added complexity of cross-layer coupling. The trilemma explains why realistic joint formulations usually simplify at least one layer, a pattern reflected in pairwise studies that often fix or aggregate the omitted third layer.

    \item \textbf{Context-dependent simplification guidance} (Sections~\ref{sec:pairwise} and~\ref{sec:future}). We show that simplifying different layers sacrifices different objectives: Planning simplification loses investment feedback, Scheduling simplification erases temporal grid and emissions dynamics, and Behavior simplification suppresses heterogeneity and equity. Thus, the choice of simplification should depend on the decision-maker's objective, with the PSB framework making the cost of each choice explicit.
\end{enumerate}

\subsection{Review scope and methodology}
This survey is a structured critical review of EV charging research, organized by decision layer, actor objective, and coupling assumption across Planning, Scheduling, and Behavior. We searched Scopus, Web of Science, ScienceDirect, IEEE Xplore, TRID, and Google Scholar using terms on infrastructure planning, charging scheduling, user behavior, and cross-layer interfaces, including siting, sizing, smart charging, dynamic pricing, vehicle-to-grid, station choice, user equilibrium, demand response, bilevel optimization, Stackelberg models, transportation-power networks, and pricing--behavior coupling.

Studies were screened through title/abstract review, full-text review, and PSB-based coding. We included papers modeling at least one PSB layer or explicitly coupling multiple layers, and coded them by layer, actor class, decision horizon, objective, method, uncertainty, grid and behavior representation, and interface variables. For pairwise studies, we also recorded which layers were endogenous, how the omitted layer was simplified, and what fidelity was lost. We excluded work centered on battery chemistry, charger hardware, converter control, or general EV adoption without direct charging-system relevance.

Section~\ref{sec:framework} introduces the PSB framework and trilemma. Sections~\ref{sec:planning}--\ref{sec:behavior} review the three layers, Section~\ref{sec:pairwise} examines pairwise coupling and omitted-layer simplifications, Section~\ref{sec:future} discusses future directions, and Section~\ref{sec:conclusion} concludes.

\section{A Three-Layer Framework for EV Charging Systems: Structure, Complexity, and Coupling}
\label{sec:framework}
This section develops the Planning--Scheduling--Behavior (PSB) framework. Section~\ref{subsec:overview} defines the three layers and their temporal scales. Section~\ref{subsec:actors} introduces the actor taxonomy. Section~\ref{subsec:cross-layer} characterizes the cross-layer coupling via interface variables. Section~\ref{subsec:trilemma} states the PSB trilemma.

\subsection{Overview of the Three-Layer Decomposition}
\label{subsec:overview}

We decompose the EV charging problem into three layers, each defined by a characteristic decision horizon and decision-maker class.

\textbf{The Planning layer (P)} concerns where to build stations, at what capacity, and how to connect them to the distribution grid. These choices are capital-intensive and difficult to reverse once construction is complete, so they fix the physical boundaries within which scheduling and user decisions must operate. Infrastructure operators, utilities, and public agencies make these decisions over a horizon of months to years.

\textbf{The Scheduling layer (S)} concerns how to allocate charging power and set prices and incentives in response to evolving vehicle arrivals and grid conditions, including day-ahead and real-time pricing. Station operators manage individual sites and decide how to split available power among connected vehicles; aggregators coordinate multiple stations or flexible loads and trade energy or ancillary services with the grid on their behalf. Both make scheduling decisions over a horizon of hours to days.

\textbf{The Behavior layer (B)} concerns how users choose where and when to charge in response to real-time signals such as prices, wait times, and station availability. Individual drivers, fleet operators, and ride-hailing drivers make these decisions over a horizon of minutes.

\subsection{Actors and Objectives}
\label{subsec:actors}
The PSB decomposition organizes the literature along the temporal-computational dimension, but a second dimension is equally important: who is making the decision at each layer. Each layer is populated by several distinct actor classes. We define an actor by the triple of decision rights, objective function, and information set. This triple, rather than the layer alone, determines which mathematical formulation to instantiate. For example, in the Planning layer, a private charge-point operator maximizes long-run profit~\cite{liu2017locating}, while a public planner maximizes spatial coverage under a budget constraint~\cite{church1974maximal}. Both use facility-location models, but their different objectives yield different optimal networks for the same candidate sites. Recognizing the implicit actor behind each formulation is therefore essential for comparing results across studies.

Table~\ref{tab:actors} lists nine actor classes across the three layers. A key pattern is that actors within the same layer can have conflicting objectives, which means that results from studies assuming different actors are not directly comparable. The implicit actor of each model reviewed in Sections~\ref{sec:planning}--\ref{sec:behavior} is flagged at the relevant subsubsection, so that the per-layer reviews can focus on model families without re-introducing the actor taxonomy.

\begin{table}[H]
\centering
%TC:ignore
\caption{Actor taxonomy across the three PSB layers, with each actor defined by its decisions and objective.}
%TC:endignore
\label{tab:actors}
\renewcommand{\arraystretch}{1.3}
\setlength{\tabcolsep}{4pt}
\begin{tabular}{@{}
  >{\centering\arraybackslash}p{0.08\linewidth}
  >{\raggedright\arraybackslash}p{0.25\linewidth}
  >{\raggedright\arraybackslash}p{0.34\linewidth}
  >{\raggedright\arraybackslash}p{0.34\linewidth}@{}}
\toprule
\textbf{Layer} & \textbf{Actor} & \textbf{Decisions} & \textbf{Objective} \\
\midrule
\multirow{3}{*}{P}
 & Charge point operator (CPO) & Siting, sizing, charger mix & Maximize net present value or long-run profit \\
\cmidrule(l){2-4}
 & Public planner & Siting, coverage targets & Maximize coverage, equity, or refuelable flow \\
\cmidrule(l){2-4}
 & Utility / distribution system operator (DSO) & Siting jointly with feeder upgrades & Minimize grid reinforcement cost \\
\midrule
\multirow{3}{*}{S}
 & Charging station operator / aggregator & Power allocation, charging price and rate setting & Maximize revenue subject to grid and quality-of-service constraints \\
\cmidrule(l){2-4}
 & Distribution system operator & Managing distribution network capacity, limiting charging load when needed & Minimize feeder overloading and voltage violations \\
\cmidrule(l){2-4}
 & Fleet manager & Routing, depot dwell, charging schedule (including individual driver assignments) & Minimize fleet operating cost under service level agreements \\
\midrule
\multirow{2}{*}{B}
 & Private driver & Station choice, timing, en-route stop, price negotiation & Maximize random utility over cost, convenience, range \\
\cmidrule(l){2-4}
 & Ride-hailing / transportation company & Charging timing, idle relocation & Maintain service availability while minimizing charging cost \\
\bottomrule
\end{tabular}
\end{table}

\subsection{Cross-Layer Coupling and Computational Structure}
\label{subsec:cross-layer}
This subsection first summarizes the mathematical structure each layer takes in isolation, then examines the three pairwise couplings and the complexity each introduces. Table~\ref{tab:interface} summarizes the six coupling edges.

\subsubsection*{Standalone Mathematical Structure}

In isolation, each layer admits a canonical formulation. Planning is a facility location problem over a finite candidate set, producing mixed-integer programs of increasing complexity as capacity, grid, and uncertainty constraints are added (Section~\ref{sec:planning})~\cite{WANG201376}. Scheduling is a sequential resource-allocation problem: in the deterministic case, the operator minimizes cost subject to energy delivery and grid limits; under stochastic arrivals and price uncertainty, this extends to a Markov decision process (MDP) (Section~\ref{sec:scheduling})~\cite{Hermans2024MPCVehicleCharging}. Behavior, once congestion feedback is introduced, becomes a fixed-point problem: each user optimizes individually given the congestion and prices implied by the current population distribution, and an equilibrium is reached when the resulting distribution reproduces itself (Section~\ref{sec:behavior}).

\subsubsection*{Planning $\times$ Scheduling}

The P$\to$S direction transmits hard capacity constraints: a station's charger count and power rating cap the dispatcher's feasible allocation. The S$\to$P return transmits observed utilization and congestion levels, informing expansion decisions~\cite{BIAN20196582}. When the two layers are coupled, the natural formulation is a bilevel program in which the planner optimizes infrastructure in the upper level, and the scheduler optimizes operations in the lower level~\cite{Luo_Pang_Ralph_1996}. Replacing the lower-level problem with its optimality conditions yields a mathematical program with equilibrium constraints (MPEC). Even with a deterministic lower level, bilevel programs with integer upper-level variables are $\Sigma_2^p$-hard in general~\cite{Luo_Pang_Ralph_1996}. When the scheduling response is modeled as an MDP rather than a convex program, its value function lacks a closed-form representation, preventing this reformulation. Most existing formulations, therefore, replace the scheduling response with a surrogate: a fixed utilization factor, a representative-day deterministic dispatch~\cite{tostadoveliz2024}, or a convex lower level amenable to optimality-condition reformulation~\cite{prakash2023bilevel}.

\subsubsection*{Scheduling $\times$ Behavior}

The S$\to$B direction transmits real-time price signals and wait times that shape user charging decisions. The B$\to$S return delivers the realized arrival rates and user-type distributions that constitute the scheduling load. When the two layers are coupled, this feedback loop transforms the single-agent scheduling problem into a multi-agent stochastic game. The resulting problem belongs to the class of equilibrium problems with equilibrium constraints (EPEC), whose feasible regions are generally nonconvex and for which constraint qualifications commonly fail~\cite{Luo_Pang_Ralph_1996}. The literature responds by splitting into two camps: dynamic-pricing papers absorb user response into a parametric demand function~\cite{obeid2023_learning_pricing_behavior}, while equilibrium papers freeze operator dispatch at its mean values~\cite{HUANG2020102179}. Each endogenizes one direction of the loop while reducing the other to a static aggregate.

\subsubsection*{Planning $\times$ Behavior}

The P$\to$B direction determines which stations are accessible to users; the B$\to$P return aggregates individual choices into long-run spatial demand patterns that shape investment returns~\cite{HUANG2020102179}. When the two layers are coupled, embedding the behavioral fixed point inside the planning bilevel program produces an MPEC whose feasible region is nonconvex due to the complementarity conditions that define the equilibrium~\cite{zhou2024_bilevel_evcs_planning}. Standard nonlinear solvers are not guaranteed to find even a local optimum, and the difficulty scales with the number of user classes and station alternatives. As a result, behavior is typically represented as a fixed or parametric demand surface that does not respond to deployment decisions over calendar time~\cite{lamontagne2023_advanced_discrete_choice}.

Taken together, the three pairwise couplings form a closed feedback system: every input that one layer takes as given is, in principle, determined endogenously by one of the other two.

\begin{table}[H]
\centering
%TC:ignore
\caption{Interface variables across the six coupling edges of the PSB framework.}
%TC:endignore
\label{tab:interface}
\renewcommand{\arraystretch}{1.25}
\setlength{\tabcolsep}{3.5pt}
\small
\begin{tabularx}{\linewidth}{@{}c X@{}}
\toprule
\textbf{Edge} & \textbf{Physical Meaning} \\
\midrule
P $\to$ S & Capacity constrains feasible scheduling domain \\
S $\to$ P & Utilization feeds back into expansion decisions \\
P $\to$ B & Siting determines reachable stations \\
B $\to$ P & Equilibrium behavior sets long-run demand \\
S $\to$ B & Price and wait time drive charging decisions \\
B $\to$ S & User types and arrival rates shape scheduling load \\
\bottomrule
\end{tabularx}
\end{table}

\subsection{The PSB Trilemma: A Fidelity-Tractability Tradeoff}
\label{subsec:trilemma}

Section~\ref{subsec:cross-layer} showed that each pairwise coupling produces a computationally hard problem class. This subsection explains why these difficulties, taken together, constitute a systematic barrier to joint modeling at realistic fidelity.

\paragraph{What is the trilemma?}
The difficulty is not that joint modeling across three layers is impossible. A small instance with five candidate sites, deterministic scheduling, and homogeneous users can be formulated and solved as a single optimization, but such instances bear little resemblance to real-world deployments involving hundreds of candidate sites, stochastic conditions, and heterogeneous user populations. The difficulty is that, at the scale and fidelity each layer requires for practically relevant applications, the problem becomes computationally intractable. The trilemma is therefore a fidelity-tractability tradeoff: practical joint formulations at a realistic scale typically reduce the modeling fidelity of at least one layer below what that layer would require if studied in isolation.

\paragraph{Why does the trilemma arise?}
The tradeoff has two sources. First, each layer in isolation is already a hard problem: facility location with integer variables is NP-hard, MDPs with large state-action spaces suffer the curse of dimensionality, and computing population equilibria scales poorly with the number of user classes and alternatives. Second, coupling any two layers compounds the difficulty: as shown in Section~\ref{subsec:cross-layer}, each pairwise coupling produces a problem class ($\Sigma_2^p$-hard bilevel programs, nonconvex EPECs, nonconvex MPECs) for which no general-purpose efficient algorithms exist~\cite{Luo_Pang_Ralph_1996}. These complexity results are not artifacts of a particular algorithm but established conclusions in the optimization and game theory literatures.

\paragraph{How hard is it to bridge the gap?}
The trilemma is not an impossibility but a tradeoff. The literature has developed approximation strategies that bridge specific pairs, but a gap remains between what these methods can handle and what real-world deployment requires. Stochastic bilevel formulations embed scheduling into the planning decision~\cite{najafi2024_stochastic_bilevel_ev}, but a realistic problem involves hundreds of candidate sites, multi-period expansion, and joint uncertainty in demand, prices, and renewable generation; current formulations achieve tractability by restricting scenarios to a handful, capturing only a fraction of real-world variability. Stackelberg game models capture leader-follower interaction between operators and users~\cite{LI2025122250,zheng2025stackelberg}, but require the follower's response to take an analytically tractable form such as a logit choice model; as Sections~\ref{subsec:richer-equilibrium} and~\ref{subsec:dynamic-behavior} show, real users satisfice, form habits, and learn from experience, none of which produce the closed-form best-response functions these formulations require. Mean-field approximations compress a large user population into a single representative agent~\cite{lasry2007mean}, but a real charging network serves commuters, ride-hailing drivers, and fleet vehicles with fundamentally different price sensitivities, time constraints, and spatial patterns; replacing them with a population average eliminates the distributional information needed to assess equity or predict group-specific responses.

This gap between method capacity and real-world complexity is reflected in the literature as a whole. The vast majority of studies model a single layer; those that jointly address two layers remain a small fraction~\cite{motlagh2026_evcs_planning_review,Motlagh2025EVCSOperationReview,SHARIATZADEH2025126167}; and formulations that integrate all three layers at realistic fidelity are essentially absent~\cite{LIANG2024124231,sun2025_bilevel_evcs_demand,cui2025multistage}. Prior surveys have noted this fragmentation~\cite{unterluggauer2022,motlagh2026_evcs_planning_review}, but they describe the symptom without diagnosing the structural cause. The trilemma provides this missing diagnosis.

\paragraph{How should the trilemma be managed?}
Different institutions control different layers, and some actors partially bridge specific pairs: a vertically integrated utility internalizes P$\leftrightarrow$S~\cite{KAZEMI20161176}, and a fleet manager internalizes S$\leftrightarrow$B~\cite{YI2021102822}. No single actor, however, spans all three layers. Joint three-layer optimization is therefore not a natural decision problem for any existing institutional role. Joint three-layer analysis, by contrast, is essential: the outcomes that matter most to regulators and planners, such as emissions, equity, and grid reliability, are emergent properties of the three-layer interaction that no single layer can predict alone. The trilemma does not excuse analysts from considering all three layers; rather, it tells them that the choice of which layer to simplify should be driven by the objective they pursue. Three decision contexts illustrate why this choice matters.

When Planning is simplified, the analyst gains a high-fidelity S$\times$B feedback loop for tariff design, but loses the ability to feed operational insights back into infrastructure decisions; if stations are in the wrong locations, no pricing strategy can compensate. When Scheduling is simplified, the analyst can jointly optimize siting and user equilibrium, but collapsing operations into a utilization factor erases the temporal structure that determines emission outcomes; a planner evaluating siting against an annual average grid may build stations that concentrate charging at peak hours, increasing emissions rather than reducing them. When Behavior is simplified, the analyst can jointly optimize siting and dispatch under grid constraints, but the model assumes users will charge as predicted regardless of network changes; the infrastructure that looks optimal on paper may perform poorly once real users interact with it. Each simplification is appropriate for some objectives and damaging for others. The trilemma thus serves not as a barrier but as a guide: it tells analysts which simplification to choose given the objective at hand, and the PSB framework makes the cost of each choice explicit.

Section~\ref{sec:pairwise} examines how each two-layer subliterature navigates this tradeoff in practice.

\section{The Planning Layer: From Basic Siting to Real-World Infrastructure Design}
\label{sec:planning}

This section reviews the Planning literature through a complexity progression: starting from the simplest siting formulation, each subsection adds one layer of real-world constraints (capacity limits, grid coupling, uncertainty, and multi-period dynamics) so that the escalation from tractable integer programs to NP-hard stochastic MILPs becomes explicit.

\subsection{Basic Siting: Where to Build}
\label{subsec:basic-siting}
The most elemental planning question is: given a set of candidate locations and a known spatial demand distribution, which sites should be activated? This is a classical facility location problem (FLP), and EV charging siting inherits its rich mathematical tradition.

Basic siting models include maximal covering~\cite{church1974maximal,WANG2025113368}, set covering~\cite{toregas1971location,YAZIR2025111786}, p-median~\cite{hakimi1964optimum,FAUSTINO2023128436,LJOSHEIM2026107289,KLOS2023103601}, p-center~\cite{hakimi1965optimum,KCHAOUBOUJELBEN2021103376,CHUNG20181,LEZCANO2025104369}, and flow-refueling formulations~\cite{kuby2005flow,GHAMAMI2016389,XIE2021117142,capar2013arc,miralinaghi2017refueling,DIMATULAC2026103970,BORLAUG2026100257}. These models differ in whether they prioritize binary coverage, average accessibility, worst-case accessibility, or path refuelability, but they share a common PSB simplification: demand is treated as fixed, station capacity is usually simplified or omitted, and both scheduling and behavioral response are externalized.

\subsubsection{Assumptions and Limitations}
The basic siting models in this subsection share several simplifying assumptions:
\begin{enumerate}
  \item \emph{No capacity constraints}: every opened station can serve unlimited demand~\cite{SPETH2025104383};
  \item \emph{Deterministic, exogenous demand}: origin-destination (OD) flows or nodal demands are fixed inputs~\cite{ANADONMARTINEZ2025104547};
  \item \emph{No grid considerations}: electrical infrastructure is implicitly assumed to be available at every candidate site~\cite{NARESHKUMAR2024114767} (relaxed in Section~\ref{subsec:real-world});
  \item \emph{Single-period}: the investment decision is made once, with no temporal staging~\cite{ye2024review}.
\end{enumerate}
These assumptions make the problem tractable as a pure integer program of moderate scale. However, each assumption will be relaxed in subsequent subsections, progressively bringing the model closer to real-world planning challenges. The first and most immediate extension is to recognize that stations have finite capacity---the subject of the next subsection.

\subsubsection{Solution Approaches}
The resulting siting models are binary integer programs that can be solved exactly using branch-and-bound or branch-and-cut for moderate-scale instances with up to a few hundred candidate sites. For larger networks, LP relaxation with rounding can provide near-optimal solutions with provable approximation guarantees~\cite{JainApproximation2001,Li7113384}, while Lagrangian relaxation can decompose MCLP and FRLM formulations by dualizing budget or coverage constraints~\cite{SONG2023119801}. When the candidate set expands to thousands of nodes, scalable greedy and interchange heuristics offer practical alternatives with empirically small optimality gaps~\cite{YIN2024133197,ZHOU2022123437}.

\subsection{Joint Siting and Sizing: How Large and What Type}
\label{subsec:siting-sizing}
In practice, the planning decision is not merely where to build but also how large and what type. Under-provisioning generates excessive waiting and degrades service quality; over-provisioning wastes capital on underutilized assets. The joint siting-and-sizing problem extends the basic formulations of Section~\ref{subsec:basic-siting} by introducing integer capacity variables that determine station size and charger mix.

\subsubsection{Capacity and Technology Decisions}

Siting-and-sizing models extend binary siting variables $x_j$ by deciding the number of chargers of each type, $n_j^{(k)}$, at station $j$, where $k \in \mathcal{K}$ captures differences in power, cost, and service capability. This allows CPOs to balance service heterogeneity and investment efficiency across Level~2 AC, DC fast charging, and related technologies~\cite{liu2017locating, Frade2011}. Some work also chooses the charging paradigm itself, including plug-in~\cite{KhanBabar02012022}, battery swapping~\cite{mak2013infrastructure}, and wireless charging~\cite{BI20181090}.

These models typically minimize site-opening and charger costs under coverage, service, and budget constraints, with linking constraints restricting chargers to opened sites. They are formulated as MILPs with linearized service constraints or MINLPs when nonlinear queueing relations are retained.

\subsubsection{Service Quality and Queueing Approximations}

With finite station capacity, demand can exceed available chargers, causing waiting or blocking. Planning models, therefore, often impose service-quality constraints on expected delay or blocking probability. A common approach embeds steady-state queueing approximations that translate charger capacity and exogenous arrival rates, usually obtained from demand forecasts~\cite{pourvaziri2024integrated} or traffic assignment, into service performance~\cite{xiao2020queueing}. Endogenizing arrivals requires coupling with the Behavior layer (Section~\ref{sec:behavior}).

The literature mainly uses three queueing paradigms. M/M/$c$ models are most common because Erlang-C yields tractable delay estimates, despite the mismatch between exponential service times and realistic CC-CV charging profiles~\cite{kinay2023charging}. Higher-fidelity M/G/$c$ approximations allow general service-time distributions but often require interpolation or simulation-based surrogates~\cite{whitt1993approximations, zhang7581114}. Erlang-B loss models apply when users are blocked rather than queued if all chargers are occupied~\cite{Bayram7060713}.

These constraints restrict feasible capacity choices and intensify tradeoffs among investment cost, coverage, waiting time, and fairness, often motivating multi-objective formulations~\cite{pourvaziri2025}. Under stochastic demand, queueing and siting-sizing decisions must be optimized jointly~\cite{kinay2023charging}, as discussed in Section~\ref{subsec:real-world}. Overall, these models add finite capacity, integer sizing, and nonlinear service constraints, while typically retaining deterministic exogenous demand, ignoring grid interactions, and using a single-period planning horizon.

\subsubsection{Solution Approaches}
The addition of integer sizing variables and nonlinear queueing constraints makes joint siting-and-sizing substantially harder than basic siting models. With linearized, piecewise-linear, or surrogate queueing expressions, the model becomes an MILP or ILP solvable for small to moderate instances~\cite{YANG2017462}. For larger instances, Benders-type decomposition is often used in EV charging location and stochastic infrastructure-expansion models by separating strategic siting decisions from routing, assignment, or scenario-dependent feasibility subproblems~\cite{kinay2023charging}. In queueing-aware models, however, nonlinear waiting-time or blocking-probability dependence on integer charger capacity makes clean decomposition difficult. Thus, metaheuristics such as genetic algorithms and particle swarm optimization are often used to retain nonlinear queueing constraints while improving scalability~\cite{MENG2025126911}. Hybrid simulation--optimization approaches further evaluate queueing performance through simulation while optimization guides the search, especially when closed-form expressions are unavailable~\cite{LIU2024123975}.

\subsection{Real-World Complications: Grid Coupling and Uncertainty}
\label{subsec:real-world}
% ---------------------------------------------------------

The siting-and-sizing models of Section~\ref{subsec:siting-sizing} assume that electrical infrastructure is always available and that all model parameters are known with certainty. In practice, neither assumption holds. This subsection addresses two qualitatively different sources of complexity: physical constraints imposed by the power grid, and informational constraints arising from uncertainty in demand, costs, and technology trajectories.

\subsubsection{Grid Constraints and Transportation--Power Coupling}

When grid constraints are modeled explicitly, the utility or DSO becomes a co-actor whose goals, such as deferring feeder upgrades or minimizing reinforcement cost, shape station feasibility. Because charging stations are high-power electrical loads, grid-aware planning becomes a coupled transportation--power network problem, where station nodes act both as transportation service facilities and distribution-network loads.

Key grid constraints include transformer capacity, feeder thermal limits modeled through linearized branch-flow approximations~\cite{baran1989network} or second-order cone relaxations~\cite{farivar2013branch}, and voltage regulation limits affected by spatial charging demand~\cite{chakraborty2024planning}. These constraints can shift siting decisions away from high-demand areas with insufficient grid capacity.

Many formulations jointly minimize station investment and grid reinforcement costs under transportation-side siting and coverage constraints and power-side flow and voltage constraints~\cite{zhang2018pev}. Coupling arises because charger count and type determine both electrical demand and service capacity. Extensions consider transformer or feeder upgrades~\cite{BIAN20196582}, as well as co-location with PV, storage, and vehicle-to-grid resources, often yielding large-scale MINLP or mixed-integer convex models.

\subsubsection{Uncertainty-Aware Optimization}

Planning models often assume known demand, costs, and system parameters, but EV adoption, charging demand, prices, policies, and technologies are uncertain. Uncertainty-aware planning is typically handled through stochastic, robust, distributionally robust, or data-driven approaches.

\paragraph{Stochastic programming}
Stochastic programming represents uncertainty with weighted scenarios and optimizes expected performance. In two-stage models, siting and sizing are decided before uncertainty is realized, while scenario-dependent performance is evaluated afterward~\cite{mirhassani2020two}. Extensions consider electric bus systems with uncertain travel times and battery degradation~\cite{zhou2023bus}. The main limitation is scenario explosion, commonly addressed by sample average approximation or scenario reduction.

\paragraph{Robust and distributionally robust optimization}
Robust optimization seeks solutions feasible for all realizations in an uncertainty set, with the set design governing conservatism and tractability~\cite{xie2018planning}. In EV charging planning, it is used to hedge against demand and renewable generation uncertainty. Distributionally robust optimization instead protects against the worst-case distribution within an ambiguity set, requiring less information than stochastic programming while reducing the conservatism of classical robustness~\cite{zhou2020planning}.

\paragraph{Data-driven and learning-based complements}
Machine learning increasingly supports uncertainty-aware planning by forecasting spatiotemporal charging demand from transactions, GPS traces, or travel surveys~\cite{WANG2023104205,JIANG2024103935,ORZECHOWSKI2023100267}, or by constructing empirically calibrated uncertainty sets~\cite{bertsimas2018data}. Recent predict-then-optimize methods align forecasting loss with downstream planning objectives~\cite{elmachtoub2022smart}. However, EV applications remain early, focusing mostly on demand forecasting for existing stations rather than new siting decisions~\cite{Li9335962,HE2024133885}.

Overall, uncertainty-aware models relax deterministic-demand assumptions but add computational burden through scenario growth, worst-case constraints, or prediction-error propagation, while often retaining simplified grid and single-period planning assumptions.

\subsubsection{Solution Approaches}
Real-world extensions require solution strategies beyond those in Sections~\ref{subsec:basic-siting} and~\ref{subsec:siting-sizing}. Grid-coupled models are typically large-scale MINLPs; SOCP relaxations of power-flow equations~\cite{farivar2013branch} can reduce them to mixed-integer SOCPs solvable for moderate networks~\cite{zhang8047328}, while larger cases often use decomposition between a transportation-side siting/sizing master problem and a power-side load-flow feasibility subproblem~\cite{zhang8384307}. ADMM has also been used to distribute computation across transportation and grid operators~\cite{SHI2025124449}. For uncertainty-aware models, SAA converts two-stage stochastic programs into deterministic MILPs using sampled scenarios~\cite{faridimehr2019stochastic,kabli2020stochastic}, scenario reduction lowers computational burden~\cite{wang2025stochasticnature}, and distributionally robust models are commonly solved by column-and-constraint generation~\cite{zhou8984288}.

\subsection{Multi-Period Dynamic Expansion Planning}
\label{subsec:multi-period}

All preceding formulations make investment decisions in a single period. In practice, EV adoption follows a gradual trajectory spanning one to two decades, and infrastructure investment must be staged accordingly: building everything at once wastes capital on underutilized assets, while building too little creates bottlenecks that suppress adoption. Multi-period planning formalizes this temporal sequencing.

\subsubsection{Staged Investment Decisions}

The planning horizon is divided into multiple periods (typically years), and the planner decides at each period which stations to open, expand, or upgrade~\cite{zhang2023multiperiod}. The objective minimizes the discounted sum of incremental investment and operational costs across all periods, extending the single-period formulation of Section~\ref{subsec:siting-sizing} with a temporal dimension and a discount rate. In the EV charging literature, multi-period models typically incorporate several domain-specific features: charger technology upgrades (e.g., replacing Level~2 chargers with DC fast chargers as demand grows), coordination with distribution network reinforcement schedules so that grid capacity is available when new stations come online~\cite{cui2025multistage}, and demand trajectories tied to regional EV adoption forecasts rather than generic growth curves~\cite{ZHANG20175Incorporating}. Some formulations also allow station relocation, relaxing the irreversibility assumption when demand patterns shift or when previously installed stations become underutilized~\cite{li2016multiperiod_relocation}.

The decisions across periods are not independent. Endogenous coupling arises from investment irreversibility (stations once built cannot be removed), monotonic capacity expansion (charger counts can only increase), technology path dependence (early charger choices constrain later upgrades), and time-phased budget constraints. Exogenous drivers include EV fleet growth, charging technology advancement, and the emergence of alternative paradigms such as battery swapping and wireless charging.

%加文献
\subsubsection{Interaction with Uncertainty}

When a multi-period structure is combined with uncertainty, the result is a multi-stage stochastic program~\cite{ye2024review}. The scenario tree branches at each decision period, and the number of scenarios grows exponentially with the number of stages, creating a curse of dimensionality in both the decision and uncertainty spaces. Unlike the single-period uncertainty models of Section~\ref{subsec:real-world}, where recourse decisions are made once after uncertainty is revealed, multi-stage formulations require a contingent plan at every period that respects non-anticipativity: the planner cannot use information from future periods when making current decisions. This temporal nesting of decisions under uncertainty is what distinguishes multi-period stochastic planning from its single-period counterpart and drives the need for specialized decomposition methods.

\subsubsection{Solution Approaches}
Even deterministic multi-period expansion planning leads to large MILPs whose size grows with periods, candidate sites, and charger types. Common solution methods include temporal decomposition and rolling-horizon heuristics, which solve truncated planning windows and fix near-term decisions before advancing; such methods have been used for fast-charging siting under evolving EV adoption and demand~\cite{anjos2020increasing}. With uncertainty, multi-stage stochastic programs become substantially harder~\cite{birge2011introduction}. Nested Benders decomposes by stage and passes feasibility/optimality cuts~\cite{KADRI2020104888}, while progressive hedging decomposes by scenario and enforces non-anticipativity, often with SAA for stochastic charging demand and expansion planning~\cite{quddus2019expansion}. SDDP can avoid explicit scenario enumeration under convex recourse by approximating cost-to-go functions~\cite{xie2018planning}, and deep reinforcement learning has recently been used to learn investment policies from simulated demand and grid trajectories~\cite{schroer2024drl}. These complexities arise within the Planning layer alone; Sections~\ref{sec:scheduling} and~\ref{sec:behavior} revisit the Scheduling and Behavior layers when infrastructure decisions are fixed.

\section{The Scheduling Layer: From Basic Dispatch to Real-Time Adaptive Control}
\label{sec:scheduling}

This section traces a complexity progression in the Scheduling literature: from deterministic offline dispatch, through bidirectional scheduling under grid constraints, to stochastic online decision-making where uncertainty and state feedback require sequential adaptive methods. Unlike the Planning layer, the Scheduling layer does not have a dedicated multi-period extension: scheduling decisions are inherently short-horizon (hours to days), so the temporal staging that drives Section~\ref{subsec:multi-period} does not arise here.

\subsection{Multi-EV Scheduling Regimes}
\label{subsec:basic-scheduling}

The basic scheduling problem asks when each connected EV should charge and at what rate under limited station capacity. This foundation of smart charging extends from deterministic offline scheduling to bidirectional operation, pricing-based coordination, and real-time adaptive control~\cite{Motlagh2025EVCSOperationReview,Sadeghian2022EVSmartChargingReview}.

Multi-EV scheduling differs by whether vehicles are centrally coordinated or uncoordinated consumers sharing capacity. In the coordinated case, charging is coupled with routing and task assignment, creating a combinatorial optimization problem. In the uncoordinated case, arrivals, dwell times, and energy needs are revealed only at plug-in, forcing operators to allocate capacity under uncertainty about future demand.

\subsubsection{Coordinated Multi-EV: Fleet Scheduling}

The coordinated regime corresponds to the fleet manager actor in Table~\ref{tab:actors}. A fleet operator jointly manages vehicle dispatch and charging for tasks such as delivery routes, bus trips, or ride-hailing assignments. Because a charging detour affects downstream service, charging slots, route choices, and task assignments must be optimized together. The canonical model is the Electric Vehicle Routing Problem (EVRP), especially the E-VRPTW, which extends the classical VRP with battery state and recharging stations~\cite{schneider2014evrptw}. Variants incorporate partial recharging~\cite{keskin2016partial,wang2023partial}, nonlinear charging curves~\cite{montoya2017nonlinear,froger2019evrp_capacitated}, and heterogeneous fleets~\cite{hiermann2016routing,wang2024heterogeneous}; see~\cite{kucukoglu2021evrp_review} for a broader review. Related electric bus models jointly optimize trip assignment and depot or network charging under capacity constraints~\cite{zhou2024bus_charging_scheduling,duan2023integrated_bus_scheduling}, with broader reviews in~\cite{perumal2022electric_bus_review}.

Here, complexity is mainly combinatorial: route choices, task assignments, and charging slots interact, while task sequences and vehicle states are largely known to the fleet operator.

\subsubsection{Uncoordinated Multi-EV: Shared-Capacity Allocation}

The uncoordinated regime corresponds to the station operator actor in Table~\ref{tab:actors}. A single operator allocates limited station power among independently arriving vehicles, each with an arrival time, departure deadline, and energy requirement. When several vehicles charge simultaneously, their profiles become coupled by the shared station-level power limit, so allocating more power to one vehicle reduces what is available to others.

This setting falls within deterministic EV charging scheduling on parallel chargers with time-window and power-limit constraints~\cite{DOLGUI2025221}. Objectives include minimizing electricity cost under time-of-use tariffs, reducing transformer peak load~\cite{Wu9185026}, and improving fairness across heterogeneous charging needs~\cite{DREUCCI2025125417}. Overall, the problem becomes a resource-constrained scheduling program whose size scales with the number of vehicles and time slots.

\subsubsection{Assumptions and Limitations}

Despite this regime-specific difficulty, both regimes above, in their baseline form, share several restrictive assumptions that the remainder of Section~\ref{sec:scheduling} progressively relaxes:
\begin{enumerate}
  \item \emph{Deterministic arrivals and departures}: every vehicle's dwell-time window and energy requirement are known at the time of scheduling, permitting the entire horizon to be optimized in a single pass.
  \item \emph{Unidirectional charging without upstream grid modeling}: vehicles can only draw power, and network-side constraints beyond the station-level limit are ignored.
  \item \emph{Offline scheduling}: the schedule is computed once before execution, with no re-optimization as conditions evolve.
\end{enumerate}
Section~\ref{subsec:bidirectional} relaxes assumptions (2) and partially (1) by introducing bidirectional power exchange and upstream grid constraints, while Section~\ref{subsec:uncertainty-scheduling} relaxes assumptions (1) and (3) by replacing deterministic inputs with stochastic arrivals, prices, and renewable generation, and by moving from offline to online, closed-loop scheduling.

\subsubsection{Solution Approaches}
The two scheduling regimes require different solution strategies. Coordinated fleet scheduling inherits the NP-hardness of EVRP/VRP, so exact branch-and-price or branch-and-cut methods are limited to moderate instances. Larger fleets typically rely on ALNS and related metaheuristics, which scale well but sacrifice optimality guarantees~\cite{schneider2014evrptw,kucukoglu2021evrp_review}; decomposition methods that separate routing and charging offer a practical middle ground. For uncoordinated shared-capacity allocation, deadline-based online rules such as EDF and LLF are simple and effective at plug-in time~\cite{Xu7431973,Chen9663226}, while LP, MILP, or QP dispatch models optimize coupled multi-EV charging with fairness-aware objectives, such as proportional allocation by remaining energy need, but must be resolved at each decision epoch~\cite{DOLGUI2025221}.

\subsection{Bidirectional Scheduling and Grid Integration}
\label{subsec:bidirectional}
% ---------------------------------------------------------

Section~\ref{subsec:basic-scheduling} treated each vehicle as a passive, unidirectional load and ignored upstream network constraints. This subsection relaxes both assumptions. Conceptually, this is also where the aggregator and DSO actors enter: bidirectional power flows and feeder constraints are precisely the dimensions in which an aggregator monetizes flexibility and a DSO enforces network limits.
\subsubsection{V2G and Bidirectional Power Allocation}

Vehicle-to-grid (V2G) expands EV scheduling from charge-only control to bidirectional charging and discharging, turning EVs from passive loads into controllable energy resources. Because discharge is constrained by battery C-rate and inverter limits, the feasible action space is asymmetric. V2G can support peak shaving, frequency regulation, and energy arbitrage~\cite{Rao2025V2GOpportunities}; frequency regulation is especially attractive due to EV batteries' fast response~\cite{LIU201946}, while arbitrage exploits time-of-use price differences~\cite{MOHAMMADIAN2025125036}.

Bidirectional operation adds trade-offs around battery degradation, conversion losses, and mobility requirements. Since aging depends on charge rate and depth of discharge~\cite{bashash2011phev}, repeated V2G cycling can accelerate capacity fade. Degradation-aware models therefore balance V2G revenue against battery health while ensuring minimum departure energy~\cite{Lu10739340}. Related vehicle-to-building settings use EVs to supply building loads under variable prices and demand charges, and have been studied with multi-agent reinforcement learning~\cite{Liu2024V2BRL}, MDP-based online search~\cite{Sen2025V2BMDP}, and neuro-symbolic MPC-learning frameworks~\cite{iccps2026_pv2b}.
\subsubsection{Grid Constraints and Multi-Resource Coordination}

When station power exchange is large relative to local distribution capacity, grid constraints such as transformer limits, feeder thermal limits, and voltage bounds become binding. Including these constraints turns station scheduling into a joint scheduling-and-power-flow problem, often using linearized power-flow models for tractability~\cite{Ahmed2026HolisticReview}. This issue is especially important in depot charging, where concentrated transit-fleet loads require joint trip assignment, charge scheduling, and distribution-network coordination~\cite{Sen2024GridAwareTransit}.

At the network level, coordinating multiple stations on the same feeder can reduce grid stress by shifting charging loads across space and time~\cite{Saber10414177,Chen10251469Distributed}. Co-located PV and stationary storage add another layer of coordination, requiring joint dispatch of EV charging/discharging, battery cycling, and curtailable solar generation under station- and grid-level constraints~\cite{HU2025125714}.

\subsubsection{Solution Approaches}
Bidirectional scheduling and grid-coupled dispatch introduce nonlinearities from battery degradation models, power flow equations, and the expanded charge-discharge action space. When degradation is modeled as a linear or piecewise-linear function of throughput and depth of discharge, the V2G scheduling problem can be formulated as an LP or MILP solvable by commercial solvers for single-station instances~\cite{Khezri10488452}. For grid-coupled formulations, linearized branch-flow approximations reduce the power flow constraints to linear or second-order cone form, enabling joint scheduling-and-power-flow optimization at moderate network scales~\cite{ZhangJian8665870}. Multi-resource coordination problems involving EV fleets, stationary storage, and local renewables are typically formulated as MILPs with time-coupled state-of-charge constraints; for larger instances, rolling-horizon and distributed optimization methods partition the problem temporally or spatially~\cite{Rehman10122724}.

\subsection{Real-World Complications: Uncertainty and Adaptive Scheduling}       
\label{subsec:uncertainty-scheduling} 

In practice, the Scheduling layer faces short-run operational uncertainty (hours to days) that unfolds while decisions are being executed, driven by stochastic arrivals, real-time prices, and intermittent renewables~\cite{Motlagh2025EVCSOperationReview}. Two paradigms address this challenge, distinguished by how uncertainty is revealed and whether closed-loop state feedback is available.
\subsubsection{Anticipative Optimization under Uncertainty}

Anticipative scheduling commits to a schedule or contingency plan at one decision point while hedging against uncertain future demand, prices, or renewable generation.

\paragraph{Stochastic programming}
Two-stage stochastic programming first sets a day-ahead schedule, then uses recourse actions such as real-time power adjustments after uncertainty is revealed~\cite{Wu2017TwoStageStochastic}. Applications include fast-charging stations with stochastic arrivals~\cite{SEDDIG2019769}, fleet depots with uncertain trip schedules~\cite{RAFIQUE2022129856}, and joint charging--energy trading under volatile wholesale prices~\cite{ZHENG2020115977}. As in planning models, scenario richness improves fidelity but increases computational burden.

\paragraph{Robust and distributionally robust optimization}
Robust optimization protects against worst-case outcomes within uncertainty sets~\cite{Kandpal2022RobustDayAhead}, while distributionally robust optimization hedges against worst-case distributions within ambiguity sets~\cite{LI2025125269}. Scheduling uncertainty sets often cover arrival intervals, price bands, or forecast errors. Applications include robust day-ahead EV aggregator scheduling~\cite{Porras9122589} and distributionally robust coordination under ambiguous renewable forecasts~\cite{BAGHERITOOKANLOU}. These methods improve feasibility guarantees but may sacrifice expected-case performance.

\subsubsection{Sequential Decision-Making under Uncertainty}

The second paradigm abandons one-shot optimization in favor of sequential decisions that respond to the evolving system state. The scheduling problem is formalized as a Markov decision process (MDP) with state $s_t$ encoding vehicle charge levels, queue occupancy, and grid signals; action $a_t$ collecting power allocation decisions; transition kernel capturing stochastic arrivals and price dynamics; and reward measuring operational objectives (cost minimization, service quality, grid stability)~\cite{Sadeghianpourhamami8727484}. The optimal policy satisfies the Bellman optimality equation, but the state space grows combinatorially with the number of vehicles, charger types, and time granularity, rendering exact dynamic programming intractable for realistic station sizes~\cite{Jin2020OptimalDeadline}. This intractability motivates the approximate solution methods discussed below.

\subsubsection{Solution Approaches}

Anticipative scheduling uses methods similar to the Planning layer (Section~\ref{subsec:real-world}), including SAA, column-and-constraint generation, and scenario reduction for uncertain arrivals and prices. Sequential control often uses MPC, which repeatedly solves a finite-horizon problem, implements the first action, and re-optimizes as new information arrives; it is attractive because it needs no offline training, handles hard constraints, and is robust to forecast errors, with applications in workplace charging~\cite{Hermans2024MPCVehicleCharging}, PV-coupled fast charging~\cite{Wamalwa2025AdaptiveMPC}, and fleet charging under time-varying prices~\cite{Skugor2025HierarchicalMPC}. When repeated optimization is too costly, or models are unavailable, DRL can learn policies from simulation for real-time scheduling~\cite{Wan8521585,Li2023LargeScaleDRL} and dynamic pricing~\cite{Cui2023DynamicPricingDRL}, though constraints are often penalty-based and generalization remains challenging. Multi-station settings extend these ideas through decentralized MARL~\cite{Yang2026CooperativeMARL} or hierarchical distributed decomposition~\cite{Khaki2019HDEVCS}. Overall, this section treats infrastructure and user behavior as exogenous, while the next section models user response to prices, wait times, and station attributes.

\section{The Behavior Layer: From Individual Choice to Dynamic Equilibrium}
\label{sec:behavior}

EV charging choice is modeled within the random utility framework~\cite{mcfadden1973conditional, ben1985discrete}, but two features distinguish it from generic facility choice. First, the feasible station set depends on the vehicle's remaining energy; second, the charging context (en route versus destination) affects both the relevant alternatives and the attributes that shape utility.

\subsection{Basic Choice: Where to Charge}
\label{subsec:basic-choice}

The most basic behavioral question is straightforward: given a set of available charging stations, which one does a user choose? As in the broader transportation demand literature, EV charging choice is commonly modeled within a random utility framework~\cite{mcfadden1973conditional, ben1985discrete}. However, two features make EV charging different from generic facility choice. First, the set of stations that a driver can realistically consider depends on the vehicle's remaining energy. Second, the charging context, especially the distinction between en-route and destination charging, affects both the relevant alternatives and the factors that shape utility.

\subsubsection{The Discrete-Choice Benchmark}

A natural benchmark in this literature is the standard discrete-choice model, which most directly describes a private driver such as a commuter or long-distance traveler. In this benchmark setting, the feasible station set is treated as given, and the user's decision depends on personal characteristics together with observable station attributes. The central idea is that each driver compares the available charging options and chooses the one with the highest perceived utility.

In practice, the systematic part of utility is typically specified as a function of variables such as distance, charging price, charger type, and posted waiting time, while the remaining term captures unobserved preference variation. Under the usual assumptions, this leads to the multinomial logit (MNL) model~\cite{mcfadden1973conditional}, which remains the canonical benchmark for charging-choice analysis. Empirical applications of MNL to EV charging station choice have identified travel distance, charging cost, and waiting time as the dominant attributes~\cite{BHAT2024177}. When stations can be grouped naturally, for example by area, operator, or charging technology, nested logit is often used to relax the independence-of-irrelevant-alternatives assumption by allowing correlation within groups~\cite{ben1985discrete}. In this section, MNL and nested logit serve as the baseline choice models; richer formulations such as mixed logit and latent-class models are discussed later in Section~\ref{subsec:richer-equilibrium}.

\subsubsection{Endogenous Feasibility and Charging Context}

EV charging choice differs from generic facility choice because the feasible choice set depends on the vehicle state. A station is relevant only if the EV has enough remaining energy to reach it, so battery state directly shapes feasible alternatives~\cite{YANG2016190Modeling}. In practice, reachability also depends on traffic, temperature, and driving style, making battery-based feasibility a first-order approximation.

Charging context further affects choice-set construction and utility design. In en-route charging, drivers must recharge to complete an ongoing trip, so route geometry, remaining range, detour distance, charger speed, and waiting time are central. In destination charging, where vehicles charge while already parked at home, work, or commercial sites, range constraints are weaker and the trade-off shifts toward cost, convenience, and dwell-time compatibility. These differences explain why empirical studies often model en-route and destination charging separately~\cite{SICA2025106175, BHAT2024177}.

\subsubsection{Assumptions and Limitations}

The basic choice models in this subsection share three simplifying assumptions:
\begin{enumerate}
  \item \emph{Exogenous and independent attributes}: station prices, wait times, and availability are fixed inputs, unaffected by the aggregate choices of other users, so that each user's utility depends only on station attributes and personal characteristics;
  \item \emph{Static}: the decision is made once, with no temporal dynamics or repeated interaction;
  \item \emph{Utility-maximizing response}: each user selects the alternative that maximizes perceived random utility, given the attributes of available alternatives.
\end{enumerate}
These assumptions reduce the problem to a standard discrete-choice model. The subsections that follow progressively relax them: first endogenizing congestion (Section~\ref{subsec:static-equilibrium}), then enriching the resulting equilibrium along several dimensions (Section~\ref{subsec:richer-equilibrium}), and finally introducing temporal dynamics (Section~\ref{subsec:dynamic-behavior}).

\subsection{Choice with Feedback: Static Equilibrium}
\label{subsec:static-equilibrium}
% ---------------------------------------------------------

The equilibrium analysis below treats users as a homogeneous population. Once assumption~(i) of Section~\ref{subsec:basic-choice} is relaxed, station attributes are no longer exogenous: the wait time at a station depends on how many users select it, which in turn depends on the wait times, creating a circular dependence that transforms the individual choice problem into a population-level consistency problem.

\subsubsection{From Individual Choice to Population Fixed Point}

At the population level, each user's station choice depends on congestion, while congestion is determined by the aggregate distribution of users across station-time-route alternatives. An equilibrium occurs when congestion-aware individual choices reproduce the same aggregate distribution. This fixed-point structure, introduced in Section~\ref{sec:framework}, has been used to model congestion-coupled EV station choice in urban networks~\cite{HUANG2020102179}.
\subsubsection{Equilibrium Benchmarks}

Two classical benchmarks are commonly used. Wardrop user equilibrium (UE) assumes deterministic utility: all used stations or routes have equal generalized cost, and no unused alternative is cheaper~\cite{wardrop1952road}. It provides the deterministic fixed-point benchmark and can also be written as a variational inequality. Stochastic user equilibrium (SUE) instead allows random utility, yielding a logit-based fixed point in which choice probabilities depend on congestion-dependent utilities~\cite{LIU2023120943}.

This section adopts SUE as the main benchmark because stochastic perception fits discrete charging choice under imperfect information. Wardrop UE is recovered when perception noise vanishes, but in EV charging it is mainly a conceptual anchor because SOC-dependent reachability and range constraints make feasible paths state-dependent~\cite{XU2024104419}. When users jointly choose routes and charging stations, the equilibrium extends traffic assignment to EV-specific and coupled transportation--power network equilibria~\cite{LI2025122250, ZENG2025104438}.

\subsubsection{Solution Approaches}

Computing the fixed-point equilibrium requires iterative methods whose choice depends on the equilibrium concept. For deterministic Wardrop UE, the equivalent convex optimization formulation~\cite{beckmann1956studies} permits gradient-based and Frank-Wolfe algorithms, with convergence guaranteed under monotone cost mappings. For stochastic user equilibrium, the method of successive averages (MSA) is the most widely adopted approach: at each iteration, a new auxiliary flow is generated from the current congestion-dependent choice probabilities and averaged with the incumbent solution~\cite{sheffi1982algorithm, powell1982convergence}. MSA is simple to implement and converges under mild conditions, but convergence can be slow for large networks. More recently, path-based and projection-based variational inequality solvers have been adapted to EV-specific network equilibrium problems where SOC-dependent feasibility constraints alter the structure of the feasible set~\cite{XU2024104419}.

\subsection{Relaxing the Benchmark Assumptions}
\label{subsec:richer-equilibrium}
% ---------------------------------------------------------

The static equilibrium of Section~\ref{subsec:static-equilibrium} rests on three benchmark assumptions: users are homogeneous, respond according to a single utility-consistent rule, and are individually negligible. This subsection relaxes each assumption along an independent dimension: user type, decision rule, and population structure. These are parallel relaxations, not a progression; each can be pursued independently of the others.
\subsubsection{User Heterogeneity and Multi-Class Equilibrium}

Charging users differ in value of time, price sensitivity, range anxiety, and vehicle attributes~\cite{thorhauge2024range, PELLEGRINI2024104722}. Multi-class equilibrium models capture this by dividing users into classes with distinct utility functions, while requiring each class's choices to be consistent with the congestion jointly generated by all classes.

This coupling increases state-space size and makes uniqueness harder to guarantee. When heterogeneity is unobserved, latent-class and mixed-logit models can infer population type distributions for use in these coupled equilibria~\cite{yang2024latentclass,zhou2025smartcharging}.

\subsubsection{Bounded Rationality}

Benchmark equilibria assume users respond optimally to the modeled utility, but real drivers may satisfice, follow habits, or perceive costs imperfectly. Bounded rationality captures these departures from utility-consistent response through several parallel mechanisms.

Relaxed optimization uses an indifference band, so users switch only when cost differences exceed a threshold~\cite{zhao2016bounded}. Switching-friction models add habit or inertia costs for changing from familiar alternatives, with the dynamic formation of habits deferred to Section~\ref{subsec:dynamic-behavior}~\cite{thorhauge2020habit}. Distorted value perception, including prospect-theory and reference-dependent utility, captures asymmetric reactions to gains and losses, such as aversion to unexpected price or wait-time increases~\cite{gao2021cpt}. Noisy-response models such as quantal response equilibrium replace exact best response with probabilistic choice whose error declines with payoff differences~\cite{zhao2014qre}; in EV charging, QRE has been combined with level-$k$ reasoning for congestion-aware station choice~\cite{Xiong2021ChargingStrategy}.

Each mechanism modifies the response map, potentially changing equilibrium existence, multiplicity, and welfare properties.

\subsubsection{Finite-Player Strategic Interaction}

This relaxation changes the population structure. When a fleet operator or aggregator controls many vehicles, each player's charging strategy can noticeably affect congestion, so the equilibrium shifts from a nonatomic fixed point to a finite-player Nash equilibrium. Strategies usually span vehicle charging schedules over time, and when congestion admits a common potential function, the equilibrium can be characterized through potential minimization~\cite{paccagnan2019}. With asymmetric roles, such as aggregators interacting with individual users or market actors, Stackelberg models capture leader--follower behavior~\cite{zheng2025stackelberg}. These relaxations enrich the static equilibrium object; the next subsection adds temporal dynamics.

\subsection{Dynamic Behavior and Learning}
\label{subsec:dynamic-behavior}

The equilibrium concepts developed in Sections~\ref{subsec:static-equilibrium} and~\ref{subsec:richer-equilibrium} characterize self-consistent population distributions but do not address how such distributions arise or evolve. This subsection introduces the temporal dimension, moving from discrete-time adaptation rules to continuous-time evolutionary dynamics, then to the mean field game equilibrium. The day-to-day learning and adaptation models below are most plausibly empirically grounded in the TNC driver class, where repeated interaction with the system and observable earnings make learning dynamics directly identifiable.

\subsubsection{Day-to-Day Adaptation}

Day-to-day models describe how users revise charging choices over discrete periods, such as daily, based on experienced wait times, price changes, and station availability. Common update rules include best-response switching, proportional shifts toward better-performing alternatives, and reinforcement-based increases in the probability of choices with favorable outcomes~\cite{ye2021dtd}.

A central question is whether these dynamics converge to the static equilibria in Sections~\ref{subsec:static-equilibrium}--\ref{subsec:richer-equilibrium}. Convergence depends on the update rule and congestion structure, with stability analysis linking dynamic adjustment to static equilibrium~\cite{bie2010stability}.
\subsubsection{Evolutionary Dynamics}

For revision rules such as imitation or payoff-proportional updating, day-to-day adaptation can be approximated by continuous-time ordinary differential equations. Replicator dynamics are a canonical example: alternatives with above-average payoff gain population share over time~\cite{WANG2013156}. This formulation is an analytically tractable limit of specific behavioral update rules, not a separate paradigm.

Its main advantage is that ODE stability tools can analyze the convergence of the adjustment process, as shown in traffic choice dynamics~\cite{SATSUKAWA2024103039}. EV charging applications remain limited, but the framework is suitable for studying how station-choice patterns evolve as EV adoption and congestion change~\cite{HUANG2022124700}.
\subsubsection{Mean Field Game Equilibrium}

Mean field games (MFGs)~\cite{lasry2007mean, huang2006large} model dynamic equilibrium in large populations of interacting agents. They couple individual optimal control, given the population distribution, with the evolution of that distribution under aggregate decisions. The resulting fixed point describes a trajectory where no individual benefits from deviating.

MFGs extend the large-population, utility-consistent benchmark of Section~\ref{subsec:static-equilibrium} to dynamic settings, but do not directly capture the bounded rationality or finite-player structure in Section~\ref{subsec:richer-equilibrium}. In EV charging, states may include battery level, location, and time of day, while controls represent charging decisions or rates~\cite{LIN2022120198}. Existence and uniqueness often rely on monotonicity of the cost coupling~\cite{lasry2007mean}.

\subsubsection{Solution Approaches}
The dynamic behavior models vary in computational complexity. Day-to-day adaptation and evolutionary dynamics are typically solved by forward simulation or ODE integration, with convergence checked numerically or through Lyapunov arguments~\cite{smith1984stability, SMITH2016132}. MFG equilibria require coupled backward--forward systems, usually solved after state-time discretization as finite-dimensional equilibrium or complementarity problems. When analytical equilibria are unavailable, agent-based simulation provides a flexible alternative for studying how charging patterns emerge, stabilize, or oscillate through repeated interaction~\cite{YI2023101949}. ABS populates a synthetic environment with boundedly rational agents using decision rules from Sections~\ref{subsec:basic-choice}--\ref{subsec:richer-equilibrium}, and simulates population-level outcomes~\cite{WARAICH201374, WOLBERTUS2021262}. Thus, ABS is not an equilibrium concept but a computational framework for richer behavioral assumptions and environments. Overall, the Behavior layer moves from individual choice to congestion-coupled equilibrium and dynamic adaptation, while treating planning and scheduling decisions as fixed inputs. The next section relaxes these assumptions to examine cross-layer interactions.

\section{Pairwise Coupling: A Structured Gap Analysis}
\label{sec:pairwise}

This section reads the three pairwise-coupling literatures (P$\times$S, S$\times$B, P$\times$B) through the trilemma of Section~\ref{subsec:trilemma}, asking for each pair what the joint literature endogenizes, what surrogate it adopts for the exogenized third layer, and what that surrogate costs in fidelity. Across all three pairs, the dominant response is to replace the exogenized layer with a static aggregate, a recurring pattern we call the static-aggregate compromise. Table \ref{tab:pairwise_gap_analysis} provides an overview for this section.

\begin{table}[H]
\centering
%TC:ignore
\caption{Summary of pairwise coupling gaps in the PSB framework.}
%TC:endignore
\label{tab:pairwise_gap_analysis}
\small
\renewcommand{\arraystretch}{1.4}
\setlength{\tabcolsep}{4pt}
\begin{tabular}{@{}
  >{\centering\arraybackslash}p{0.12\linewidth}
  >{\raggedright\arraybackslash}p{0.2\linewidth}
  >{\raggedright\arraybackslash}p{0.25\linewidth}
  >{\raggedright\arraybackslash}p{0.25\linewidth}
  >{\raggedright\arraybackslash}p{0.25\linewidth}@{}}
\toprule
\textbf{Coupling} &
\textbf{Endogenized layers} &
\textbf{Typical omitted-layer surrogate} &
\textbf{What is lost} &
\textbf{Suitable objective} \\
\midrule
P$\times$S &
Planning and scheduling &
Behavior represented as fixed demand, fixed arrival rates, or aggregate utilization &
User heterogeneity, station choice response, adoption feedback, and equity effects &
Grid-aware infrastructure planning and capacity sizing \\
\addlinespace
S$\times$B &
Scheduling and behavior &
Planning represented as fixed station locations, fixed charger capacities, or static network topology &
Long-run investment feedback, capacity expansion effects, and siting adaptation &
Dynamic pricing, demand response, and smart charging control \\
\addlinespace
P$\times$B &
Planning and behavior &
Scheduling represented as fixed service rates, queueing functions, utilization factors, or average congestion &
Temporal grid interaction, peak-load dynamics, marginal emissions, and operational flexibility &
Access-oriented siting, equity analysis, and station-choice-aware planning \\
\bottomrule
\end{tabular}
\end{table}

\subsection{Planning $\times$ Scheduling}
\label{subsec:PS}
% ---------------------------------------------------------

As Section~\ref{subsec:cross-layer} showed, coupling planning and scheduling produces a bilevel program whose lower level must be replaced with a tractable surrogate. Across the joint P+S literature, the surrogates fall into three families arranged along a fidelity spectrum~\cite{unterluggauer2022}. Notably, the advanced scheduling methods reviewed in Section~\ref{subsec:uncertainty-scheduling} (MPC, stochastic programming, deep reinforcement learning) do not yet appear as the lower level in any mainstream joint P+S formulation.

At the lowest fidelity, the scheduling layer is collapsed into an average utilization factor fed back into the planning objective~\cite{motlagh2026_evcs_planning_review,YI2022103264,YANG2017462}. This yields a closed-form objective solvable as a standard mixed-integer program, but erases all temporal structure: peak-shaving, time-of-use response, and every feature that distinguishes one scheduling policy from another reduces to a single scalar.

At intermediate fidelity, a representative-day deterministic dispatch program is embedded as the lower level, typically a linear program over a small number of typical days~\cite{tostadoveliz2024,ZHANG2025101946}. This preserves intra-day load shape and price structure while staying within convex tractability, but flattens price volatility, ramping constraints, and V2G arbitrage into a deterministic schedule.

At the highest fidelity currently available, a full bilevel program is retained with the lower level restricted to a problem class that admits closed-form reformulation, such as a linear or convex quadratic dispatch~\cite{prakash2023bilevel}, or a tractable demand-allocation problem solved through cooperative game conditions~\cite{sun2025_bilevel_evcs_demand}. These preserve lower-level optimality but cannot accommodate stochastic arrivals or learned policies. A small frontier pushes beyond this through stochastic bilevel formulations~\cite{najafi2024_stochastic_bilevel_ev} and learning-based approximations that replace analytical lower-level reformulation with simulation-trained operational policies or configuration policies~\cite{schroer2024drl}.

The three surrogates differ in how much intra-day structure they retain, but share a common cost: each collapses the hour-scale scheduling horizon into the year-scale planning horizon, aggregating away temporal variation. This timescale collapse returns in sharper form in Section~\ref{subsec:PB}, where the gap between planning and behavioral response is even wider.

\subsection{Scheduling $\times$ Behavior}
\label{subsec:SB}

As Section~\ref{subsec:cross-layer} showed, closing the operator-user feedback loop transforms the scheduling problem into a multi-agent stochastic game whose complexity grows with the number of user types. The joint S$\times$B literature responds by splitting into two camps, each closing only one direction of the loop while reducing the other to a static aggregate.

On one side, dynamic-pricing and demand-response papers treat the operator's decisions as endogenous and absorb user response into a parametric demand function, such as a logit choice probability or a linear price elasticity~\cite{obeid2023_learning_pricing_behavior,ZHOU2025104358,KAZEMTARGHI2024109946,Xiong2017Pricing}. This reduces the operator's problem to a Stackelberg or single-level optimization, but compresses user heterogeneity and strategic interaction into a single demand curve. Rather than compressing user response into a demand curve, negotiation-based frameworks offer drivers incentive-backed options for flexibility in departure time or requested state of charge, aligning operator and user objectives through mechanism design~\cite{sen2026consent}.

On the other side, equilibrium routing and station-choice papers endogenize user response through a discrete-choice or user-equilibrium model and let operator scheduling enter only through a fixed congestion or queueing function~\cite{HUANG2020102179,CHEN2023104305,YANG2018189userchoice}. The cost is symmetric: operator pricing and dispatch are frozen, suppressing feedback from real-time adaptation into the user equilibrium.

The two camps are mirror images: each endogenizes one direction of the S$\leftrightarrow$B interface and reduces the other to a static aggregate. A small frontier of mean-field game and multi-agent reinforcement learning work attempts to close both directions simultaneously~\cite{LIN2022120198,liu2023_marl_mobile_charging,QIU2022118790}, but it remains small and mostly simulation-based. Across both camps and the frontier, the common cost is that the full bidirectional feedback between operator adaptation and user response has not yet been captured in a single tractable formulation.

\subsection{Planning $\times$ Behavior}
\label{subsec:PB}

Of the three pairs, P$\times$B has the smallest joint literature. The structural obstacle is that embedding the behavioral fixed point inside a planning bilevel program produces a mathematical program with equilibrium constraints (MPEC) whose nonconvexity admits no clean reformulation under logit choice or discrete travel decisions~\cite{zhou2024_bilevel_evcs_planning,QIAO2023104678,Xiong2017CSPP}. Two compounding reasons beyond this mathematical difficulty explain why the joint literature remains especially thin.

First, the time-scale gap between planning (years) and behavioral response (days) is on the order of $365{:}1$, far wider than the $24{:}1$ ratio in P$\times$S, making representative-day aggregation insufficient for capturing behavioral evolution over a changing network. Second, no institutional actor spans both long-horizon deployment and individual driver behavior: the P$\leftrightarrow$B edge has no institutional actor with direct control over both layers, unlike P$\leftrightarrow$S (integrated utility) or S$\leftrightarrow$B (fleet manager). The rare P$\times$B papers accordingly represent behavior as a fixed demand surface~\cite{lamontagne2023_advanced_discrete_choice}. Together, these reasons compound the mathematical obstacle and keep the joint literature thin.

\medskip

The preceding subsections document the specific forms the static-aggregate compromise takes across all three pairs. A consistent pattern emerges: the pairs whose coupling edge is spanned by an institutional actor (P$\times$S by the integrated utility, S$\times$B by the fleet manager) have generated learning-based or simulation-based alternatives to the static default, while P$\times$B, which lacks such an actor, also lacks such alternatives. Section~\ref{sec:future} uses these findings to ask a forward-looking question: given that some simplification is unavoidable, which layer should be simplified, and what does that choice cost?

\section{Future Research Directions}
\label{sec:future}
% =====================================================

Section~\ref{subsec:trilemma} showed that practical joint formulations at a realistic scale typically simplify at least one layer, and that each simplification forecloses a different set of objectives. This section outlines four directions for future research, spanning methodology, data, emerging technologies, and policy. The purpose of future work should not be to eliminate simplification entirely, but to make simplification measurable, objective-dependent, and institutionally transparent. 

\paragraph{Quantifying the cost of simplification}
The central open question is not which surrogate to use but how much each one costs in decision quality. Answering this requires controlled experiments in which the same planning or policy problem is solved at multiple surrogate fidelity levels and the outcomes are compared on decision-relevant metrics: total system cost, realized emissions, equity of access, or grid constraint violations. Such experiments would reveal, for instance, whether a representative-day scheduling surrogate preserves enough temporal variation to rank siting alternatives correctly for emission objectives, or whether a parametric demand function produces station placements that perform well once real users interact with the network. Existing work on bilevel approximation~\cite{najafi2024_stochastic_bilevel_ev} and mean-field behavior models~\cite{lasry2007mean} provides candidate surrogates at different fidelity levels, but head-to-head comparisons on a common instance are absent. Establishing these comparisons would convert the trilemma from a qualitative diagnostic into a quantitative design tool.

\paragraph{Data infrastructure and empirical calibration}
Surrogate comparison requires shared data, but the field currently lacks standardized benchmarks. Three types of data are needed in combination: charging-session records with temporal and spatial resolution (session start, duration, energy delivered, station identity), distribution feeder models with transformer ratings and voltage constraints, and disaggregate user-behavior data linking charging choices to trip purpose, income, and vehicle type. Some of these are becoming partially available through open charging-network APIs and utility smart-meter programs, but they remain fragmented across operators and jurisdictions, often proprietary, and rarely linked to behavioral covariates. On the empirical side, the behavioral models reviewed in Section~\ref{sec:behavior} offer rich theoretical structures for heterogeneity, bounded rationality, and dynamic learning, but calibration against observed charging behavior at the urban scale is rare. Bridging this gap would ground the surrogate comparison of the previous paragraph in empirically validated models rather than stylized assumptions.

\paragraph{Emerging technologies and evolving coupling structures}
The preceding two directions address how to use the current PSB framework more rigorously. The remaining two ask whether the framework itself will need to evolve. The PSB framework is built on the current paradigm of human-driven EVs charging at fixed stations. Several emerging technologies could change which coupling edges matter and which simplifications are appropriate. Autonomous vehicles shift charging decisions from individual drivers to fleet-level optimization algorithms, potentially collapsing the Behavior layer into the Scheduling layer and dissolving the S$\times$B boundary. Battery swapping replaces continuous charging with discrete swap events, changing the queueing and scheduling structure at the operational level. Dynamic wireless charging embeds infrastructure in the road surface, transforming the Planning problem from station siting to corridor design. Vehicle-to-grid and vehicle-to-building bidirectional energy exchange strengthens the coupling between scheduling and grid or building operations, making the S$\times$P interface more consequential. Each of these technologies does not simply add complexity to the existing framework; it restructures the trilemma by altering which layers are coupled and how tightly. Extending the PSB framework to accommodate these shifts is a necessary step toward keeping the diagnostic current as the technology landscape evolves.

\paragraph{Policy and institutional design}
The trilemma shapes not only modeling choices but also the policy decisions that determine who plans, operates, and regulates charging infrastructure. Consider two concrete policy scenarios. When a government decides whether to allow vertical integration between network planning and station operation, the trilemma clarifies the stakes: vertical integration internalizes the P$\leftrightarrow$S coupling, reducing the need for scheduling surrogates in planning models, but it concentrates market power and may weaken the competitive signals that drive operational efficiency. When a regulator mandates open data sharing between operators and municipal planners, it enables the planner to use higher-fidelity scheduling and behavioral inputs without controlling those layers directly, partially relieving the trilemma through institutional design rather than algorithmic improvement. More broadly, different policy objectives demand different modeling fidelity: tariff regulation requires detailed S$\times$B modeling, subsidy allocation for station construction requires P$\times$B modeling, and grid-integration standards require P$\times$S modeling. The trilemma tells each policy-maker which layer their model simplifies and, therefore, which outcomes their policy is blind to. Developing this connection between computational tradeoffs and governance design is a direction that sits squarely within the scope of transport policy research.

\section{Conclusion}
\label{sec:conclusion}
This survey introduced a three-layer Planning, Scheduling, and Behavior (PSB) framework that separates long-term infrastructure decisions, short-term operational control, and user charging response while making their coupling explicit. It also identifies a PSB trilemma: realistic fidelity across all three layers is difficult to achieve without losing tractability. Existing cross-layer studies therefore typically integrate two layers while simplifying the third through fixed assumptions or aggregate surrogates.

The central implication is that simplification is often necessary but should be deliberate. Simplifying Planning preserves operational and behavioral detail but ignores investment adaptation; simplifying Scheduling enables siting and equilibrium analysis but omits temporal grid dynamics; and simplifying Behavior supports grid-aware planning or dispatch but suppresses heterogeneity, strategic response, and equity-relevant differences. The appropriate abstraction therefore depends on the decision-maker, objective, and institutional context.

Important gaps remain. High-fidelity pairwise models are still limited, full three-layer formulations are rare, and empirical validation is weak because few benchmarks combine charging sessions, distribution-network data, and disaggregate behavioral information. User heterogeneity is often assumed rather than calibrated, and the decision-quality loss from simplified surrogates is seldom measured. Thus, it remains unclear when simplified models are adequate for investment, pricing, emissions, or equity analysis.

The PSB framework should therefore serve as a design tool: it helps researchers state which layers are modeled, which are simplified, which coupling edges are retained, and which objectives may be distorted. Future progress requires stronger layer-specific methods, benchmark datasets for surrogate evaluation, empirical calibration of charging behavior, and extensions to autonomous fleets, battery swapping, dynamic wireless charging, V2G/V2B integration, and institutional mechanisms for data sharing and coordinated decision rights.

%TC:ignore
\section*{Declaration of generative AI and AI-assisted technologies in the manuscript preparation process}

During the preparation of this work, the authors used ChatGPT to support language editing, rephrasing, and clarity of presentation. After using this tool, the authors reviewed and edited the content as needed and take full responsibility for the content of the manuscript.

\section*{Data availability}

No data were used for the research described in the article.

\bibliography{references}
%TC:endignore

\end{document}